\documentclass[11pt,a4paper]{article}
\pdfoutput=1
\usepackage{jcappub}
\def\lsim{\raise0.3ex\hbox{$\;<$\kern-0.75em\raise-1.1ex\hbox{$\sim\;$}}}
\def\gsim{\raise0.3ex\hbox{$\;>$\kern-0.75em\raise-1.1ex\hbox{$\sim\;$}}}

 \normalsize
\newcommand{\bmat}{\left(\begin{array}}
\newcommand{\emat}{\end{array}\right)}
\newcommand{\be}{\begin{equation}}
\newcommand{\ee}{\end{equation}}
\newcommand{\bea}{\begin{eqnarray}}
\newcommand{\eea}{\end{eqnarray}}

\newcommand{\n}{\nonumber\\}

\begin{document}
\begin{flushright}
CPHT-RR017.0515
\end{flushright}

\title{Single Field Inflation in Supergravity with a $U(1)$ Gauge Symmetry}
\author{L. Heurtier$^{1}$, S. Khalil$^{2,3}$ , A. Moursy$^{2}$ }
\affiliation{
$^1$Centre de Physique Th\'eorique, \'Ecole Polytechnique, CNRS, 91128 Palaiseau, France.\\
$^2$Center for Fundamental Physics, Zewail City of Science and Technology, 6 October City, Cairo, Egypt.\\
$^3$Department of Mathematics, Faculty of Science,  Ain Shams
University, Cairo, 11566, Egypt.}

\date{\today}

\abstract{
A single field inflation based on a supergravity model with  a shift symmetry and $U(1)$ extension of the MSSM is analyzed.  We show that one of the real components of the two $U(1)$ charged scalar fields plays the role of inflaton with an effective scalar potential similar to the ``new chaotic inflation'' scenario. Both non-anomalous and anomalous (with Fayet-Iliopoulos term) $U(1)$ are studied.   We show that the non-anomalous $U(1)$ scenario is consistent with data of the cosmic microwave background and recent astrophysical measurements. A possible kinetic mixing between $U(1)$  and $U(1)_{B-L}$ is considered in order to allow for natural decay channels of the inflaton, leading to a reheating epoch. Upper limits on the reheating temperature thus turn out to favour an intermediate ($\sim {\cal O}(10^{13})$ GeV) scale $B-L$ symmetry breaking.}
\maketitle

\section{Introduction}
Inflation offers a natural framework to alleviate the problems associated with the original Big Bang model, such as flatness, horizon, and monopole problem.  For more than three decades, the measurements of the cosmic microwave background (CMB) temperature
anisotropies have confirmed the predictions of inflation. In order to construct a consistent model of inflation, an extension of the Standard Model (SM) is required.  A supersymmetric extension of the SM with  Fayet-Iliopoulos (FI) $U(1)$ provides a simple realization of D-term inflation, which avoids the slow-roll problem in supergravity (known also as $\eta$-problem) \cite{Binetruy:1996xj, Halyo:1996pp}. This problem emerges due to the fact that in supergravity, the F-term SUSY breaking is universally mediated by gravity and all scalar fields, including inflaton, acquire masses of order the soft SUSY breaking masses, which are of order the Hubble parameter, hence spoiling the slow-rolling.

With minimal K\"ahler potential and superpotential of the form $W= \lambda S \phi_+ \phi_-$, a D-term hybrid inflation (DHI) is obtained with constant tree level potential $V= \frac{g^2}{2} \xi^2$, where $g$ is the gauge coupling and $\xi$ is a FI D-term.  Such non-zero energy density breaks SUSY and non-trivial radiative corrections are obtained, which lift the flatness of this trajectory and produce the necessary slope for driving inflation. However, hybrid inflation predicts spectral index close to one \cite{Dvali:1994ms}, which is not consistent with the recent observations \cite{Hinshaw:2012aka, Ade:2013zuv}.  On the other hand, chaotic inflation induced by $F$-terms in supergravity has been explored in Ref.\cite{Kawasaki:2000yn, Yamaguchi:2000vm, Brax:2005jv} by introducing a shift symmetry. Therefore, the K\"{a}hler potential depends only on the imaginary part of a scalar field. Thus, the real part of this field (which plays the role of inflaton) does not suffer from exponential growth. In addition, it was emphasized that a stabilizer field, which has no shift symmetry, is also required so that the inflationary potential can be bounded from below.  The stabilizer field can be interpreted in string theory as a three-form multiplet \cite{Dudas:2014pva}.

It is tempting to use a singlet scalar field as inflaton so that one can impose the shift symmetry as mentioned above. However, considering scalar fields which transform non-trivially under a $U(1)$ gauge group (or other gauge symmetry) is preferred to dilute any topological defect associated with the breaking of this symmetry \cite{Jeannerot:2000sv}.
In addition, it allows for natural couplings between these fields and the SM particles as well as right-handed neutrinos, that may give rise to reheating process after the inflation ends \cite{Jeannerot:2002wt}.

In this paper we analyse inflation in a supergravity framework  {of} $U(1)$  {gauge} symmetry with and without FI term. We introduce two fields, $\phi_1$ and $\phi_2$, carrying opposite charges under a $U(1)$ symmetry, to which we provide a shift symmetry in the K\"ahler in addition to a singlet scalar field $S$ as stabilizer. We show that a linear combination of the real parts of $\phi_1$ and $\phi_2$ will play the role of inflaton and a single field inflation potential, similar to the "new" chaotic inflation potential \cite{Kallosh:2010ug} is obtained. Our approach is somehow similar to \cite{Antusch:2010va}, although our setup has the advantage to be rather minimal and an explicit formulation in SUGRA is provided.

The paper is organized as follows. In Section 2 we study the inflation in shifted symmetric $U(1)$ SUSY model. We show that large (super-Planckian) field is required for consistent inflation.  In Section 3 we consider the possibility of a non vanishing FI term and study its influence on the inflationary dynamics. Section 3 is devoted for analysing the reheating process after the end of inflation through the decay of inflaton to two right-handed neutrinos. Our conclusions are given in Section 4.
%
\section{Shift Symmetric SUSY model} \label{sec:model1}
We are interested in building  {an inflationary model}  {with the inflaton} field charged under a $U(1)$ symmetry. To release such a scenario in a supergravity framework it is well known that one can impose a shift symmetry to circumvent the so called $\eta$-problem of seeing the inflaton appear in the K\"ahler potential,  which makes the inflaton potential much too steep \cite{Copeland:1994vg, Dine:1995uk}. Yet, defining a shift symmetry in the case where the inflaton carries some charge is a bit more involved than it is in the case of a singlet inflaton\footnote{Attempts using multi-field approach could maybe solve the $\eta$-problem \cite{add2} in a different way, but stand outside of from the framework of this study.}. Let us introduce this question by starting with the most minimal material, that is two superfields $\phi_1$ and $\phi_2$ carrying opposite charges under a $U(1)$ symmetry. The two complex scalar components contain thus four real ones among which one could potentially drive inflation. A possible choice for realizing a shift symmetry is to impose the following invariance in the K\"ahler potential.
\bea
\phi_1 \,&\rightarrow & \,\phi_1+i c \nonumber\\
\phi_2 \,&\rightarrow& \,\phi_2+i c \nonumber\\
\eea
which can arise by defining a K\"ahler potential of the sort
\bea\label{KPot1}
K= |\phi_1+\bar{\phi_2}|^2\,.
\eea
Along this punchline, and in order to get a single field inflation scenario as an effective theory, one needs to make fields that are not the inflation get masses higher than the Hubble scale during inflation. As mentioned in the introduction, a singlet {\em stabilizer} field $S$ will play an important role in providing heavy masses to those degrees of freedom \cite{Kallosh:2010ug, Kallosh:2010xz}. Let us hence start with the following $\mathcal N=1$ supergravity model whose K\"{a}hler potential is
\bea\label{KPot1}
K= |\phi_1+\bar{\phi_2}|^2+|S|^2-\zeta|S|^4,
\eea
The additional quartic term in $S$ is added in order to make the latter sufficiently heavy so it does not perturb the inflation dynamics as well.

In this respect, the most general renormalizable superpotential preserving R-symmetry is given by
\bea\label{SPot1}
W=\lambda  S(\phi_1 \phi_2+M^2),\eea
where $R[S]=2$ and $R[\phi_1]=-R[\phi_2]$ and $M$ is some dimensionful mass parameter. Note that we will work in what follows with masses in Planck units, by setting the reduced Planck mass to unity. Although the chosen superpotential is here somehow similar to the one exposed in models dealing with SUSY GUTs or FI terms hybrid inflation \cite{Dvali:1994ms, Wieck:2014xxa}, in which a neutral inflaton interact with charged superfields called the waterfall fields, we stress the fact that contrarily to this kind of models, our inflaton field will be here explicitly charged under the $U(1)$ symmetry.

For now, we do not consider any Fayet-Iliopoulos term.  {Therefore the D-term potential} is given by
\begin{equation}\label{VD1}
V_D=\frac{g^2}{2} \Big(|\phi_1|^2-|\phi_2|^2\Big)^2\,,
\end{equation}
and the  {F-term potential} is, in units where the reduced Planck mass is set to 1,
\begin{equation}
V_F=e^K\left[K^{I \bar J}D_{I}W \overline{D_{\bar J}W}-3|W|^2\right]\,,
\end{equation}
where the index $\alpha$ runs over the superfields $\{\phi_1,\phi_2,S\}$ and $D_{I}W=\partial_{I}W+W \partial_{I}K$.
Accordingly $V_F$ will take the form
\bea\label{GeneralPotential}
V_F&=& e^K\Bigg[\left((1-4\zeta |S|^2)^{-1}\Big[1+|S|^2(1-2\zeta |S|^2)\Big]\right)^{2}|\phi_1 \phi_2+M^2|^2 \\ \nonumber&&
+|S|^2\left(\Big|\phi_1+(\phi_1 \phi_2+M^2)(\phi_1+\bar{\phi_2})\Big|^2+\Big|\phi_2+(\phi_1 \phi_2+M^2)(\bar{\phi_1} +\phi_2)\Big|^2\right)\\ \nonumber&&
-3|S|^2|\phi_1 \phi_2+M^2|^2\Bigg]\,,
\eea
It turns out that a supersymmetric minimum for the potential $V= V_F+V_D$ is located at
\bea
\langle S\rangle =0 \,, \hspace{0.5 cm} \langle \phi_1 \phi_2 \rangle = -M^2\hspace{0.5cm}\text{and}\hspace{0.5cm}|\phi_1|^2=|\phi_2|^2=M^2\,.
\eea
 {Since} the K\"{a}hler potential (\ref{KPot1}) {depends} only on the combination $\phi_1+ \bar\phi_2$, it is convenient to express the potential in terms of the combinations $\phi_1 \pm \bar \phi_2$ and the supersymmetric minimum becomes
\begin{equation}\label{newvacuum}
\langle S\rangle =\langle \phi_1 + \bar \phi_2\rangle = 0 \,, \hspace{0.5 cm} | \phi_1 -\bar \phi_2 |^2 = 4 M^2\,.
\end{equation}
Using this redefinition, {the} complex fields can be written {as}
\bea\label{def}
S&\equiv& s+i \sigma \\ \nonumber
\phi_1+\bar \phi_2 &\equiv& \alpha+i \beta \\ \nonumber
\phi_1-\bar \phi_2 &\equiv& \rho \ e^{i \theta / 2M} \,,
\eea
and the minimum (\ref{newvacuum}) is then given by
\bea\label{vac}
\langle s\rangle =\langle \sigma\rangle  =\langle \alpha\rangle =\langle \beta\rangle =0
\hspace{0.5 cm}\text{and} \hspace{0.5 cm} \langle \rho \rangle = 2M\,.
\eea
The supersymmetric vacuum thus presents a $U(1)$ symmetry, which is let invariant by the massless goldstone boson $\theta$ as we will see. Furthermore, the K\"{a}hler potential~(\ref{KPot1}) is independent on the fields $\rho$ and $\theta$. In other words, the latter fields are two potential candidates for driving inflation during a 60 e-folds slow roll regime. However one has to know what is the mass range of these fields in order to determine which fields quantum fluctuations will dominate the energy density during the inflationary period.

The tree level masses are given in the true vacuum (\ref{vac}) by
\begin{eqnarray}
 m_s^2&=&m_{\sigma}^2=4M^2\lambda^2\,,\\
 m_{\rho}^2&=&m_{\beta}^2=2M^2\lambda^2\,,\\
 m_{\alpha}^2&=&4g^2M^2\,,~~m_{\theta}^2=0\,.
 \end{eqnarray}
As mentioned above, the field $\theta$ thus appears to be exactly massless in the vacuum. The latter turns out to be the Nambu Goldstone boson arising after spontaneous breaking of the $U(1)$ gauge symmetry. Indeed, the gauge lagrangian
\begin{equation}\label{gaugelag}
 \mathcal L_G\equiv K_{I\bar J}D_{\mu}\phi^I \overline{D^{\mu}\phi^{J}} = |(\partial_{\mu}-igA_{\mu})\phi_1|^2+|(\partial_{\mu}+igA_{\mu})\phi_2|^2\,,
\end{equation}
after integrating out massive fields contains a mix term
\begin{equation}
 -2 M g A^{\mu}\partial_{\mu}\theta\,,
\end{equation}
which can be absorbed in the gauge field by a unitary transformation, see Appendix~\ref{unitarygauge}.

\begin{figure}
 \begin{center}
  \includegraphics[width=0.7\linewidth]{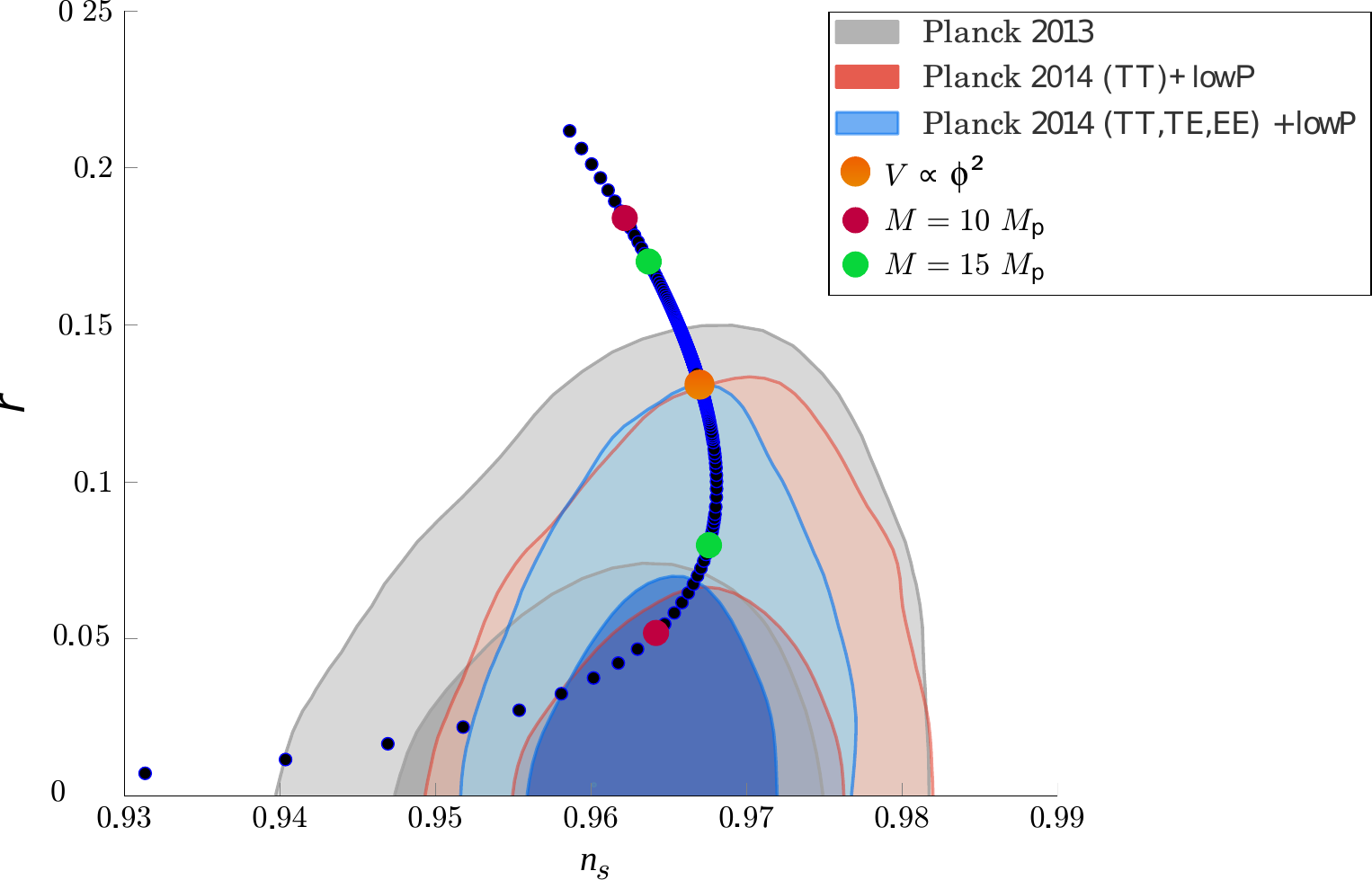}
  \caption{\label{obs_Planck}{\footnotesize Observables arising from 60 e-folds of inflation using the potential $V=\frac{\lambda ^2}{16} \left(4 M^2- \rho^2\right)^2$ in the case where $M \gg M_p$. The orange dot stands for the chaotic case. Upper part of the curve is released while the inflaton starts from high initial values of the inflaton vev whereas the lower part depicts observables in the case where the inflaton starts from low values. Purple and green dots stand respectively for the cases where $M =10\ M_p$ and $M =15\ M_p$.} }
 \end{center}
\end{figure}

The field $\theta$ turns hence out to be an unphysical degree of freedom. The only physical field not appearing in the K\"ahler is thus $\rho$ which will be from now on our inflaton candidate. So that Inflation can last for 60 e-folds, $\rho$ is assumed to take large values at initial time of the slow roll period. Other fields being present in the K\"ahler will shortly stabilize to zero and the masses turn out to be
\begin{eqnarray}\label{masses}
 (m^{\text{inf}}_s)^2&=&(m^{^{\text{inf}}}_{\sigma})^2=\frac{\lambda ^2}{2}  \left(2 \rho^2+\zeta  \left(\rho^2-4 M^2\right)^2\right) \,,\\
 (m^{^{\text{inf}}}_{\alpha})^2&=&g^2 \rho^2 + \frac{\lambda ^2}{8}(\rho^2-4 M^2) \left(\rho^2-4 M^2-2\right)\,,\\
 (m^{^{\text{inf}}}_{\beta})^2&=&\frac{\lambda ^2}{8} \left(\rho^4-8 M^2\left(\rho^2-1\right) + 2 \rho^2+16 M^4\right)\,,\\
 (m^{^{\text{inf}}}_{\rho})^2&=&\frac{\lambda^2}{4}(3\rho^2-4M^2)\,,\\
 (m^{^{\text{inf}}}_{\theta})^2&=&0\,.
 \end{eqnarray}
Again, the real field $\theta$ is massless and could in principle,  being the lightest field involved in the inflationary dynamics, perturb inflation. However we show in appendix~\ref{unitarygauge} that the latter remains absorbable by the gauge field -- and is thus an unphysical degree of freedom -- during the whole Inflation period. Therefore, from now on, we consider the inflaton to be the single field $\rho$. The inflationary trajectory hence
corresponds to\footnote{Note that for large values of the inflaton, a second minimum exists for $\alpha_1$ at $\sim \rho$ which disappears while $\rho$ approaches its minimum. Such values of $\alpha$ however generate large off diagonal terms in the mass matrix which destroy the inflation scenario. We thus assume in the following that $\alpha$ starts at small values and consequently stays stabilized at zero.} $S=\alpha=\beta=0$.
For this hypothesis to be valid, one yet  {needs} to check that $\rho$ is the lightest field involved in the scenario and that all the other fields stay heavier than the Hubble scale during inflation. The Hubble parameter during inflation is given by
\bea\label{H1}
H^2(\rho)=\frac{\lambda ^2}{48} \left(4 M^2- \rho^2\right)^2\,.
\eea
One should first note that the inflaton mass $m_{\rho}^2 \sim \lambda^2 M^2$ can be made lighter than the Hubble scale $H^2\sim \lambda^2 M^4$ only in the case where $M\gg 1$
in Planck units. We will assume this in the rest of the paper, ensuring that the single field inflation scenario is valid. Accordingly, the field dependent tree-level masses of the fields (\ref{masses}) can be expressed as
\begin{eqnarray}\label{massesH}
 (m^{\text{inf}}_s)^2&=&(m^{^{\text{inf}}}_{\sigma})^2=\lambda ^2 \rho^2+ 24\ \zeta H^2\,,\\
 (m^{^{\text{inf}}}_{\alpha})^2&=&g^2 \rho^2 + 6H^2-\frac{\lambda^2}{4}(\rho-4M^2)\,,\\
 (m^{^{\text{inf}}}_{\beta})^2&=&12\ H^2+\frac{\lambda ^2}{4} \left( 4M^2 + \rho^2\right)\,.\\
 (m^{^{\text{inf}}}_{\rho})^2&=&\frac{\lambda^2}{4}(3\rho^2-4M^2)\,,
 \end{eqnarray}
 where all the spectator fields are indeed ensured to be heavier than the Hubble scale if $\zeta \gtrsim 1/24$.  {Therefore, setting the heavy fields to their minima in the F-term and D-term parts of the scalar potential will result the single field inflationary potential}
\bea\label{Vinf1}
V_{\rm inf}(\rho)=\frac{\lambda ^2}{16} \left(4 M^2- \rho^2\right)^2\,.
\eea
Such a potential has already been studied in \cite{Kawasaki:2001as, Kallosh:2010ug, Kallosh:2007wm} in details. Small values of $M$ make the potential essentially quartic and is ruled out by WMAP measurements, hypothesis which is anyway excluded as mentioned above by the requirement of single field inflation in our model. The case of large $M\gg 1$ is opening two different possible scenarios of viable quadratic inflation\footnote{Note that similar {\em mexican hat}-like inflationary potential can be obtained in Coleman-Weinberg potentials \cite{add1} leading to similar observables.}. The case where $\rho$ starts from small values is similar to the {\em natural inflation} picture while the case where $\rho$ starts from large initial conditions is the ``new'' chaotic inflation scenario with mass $m_{\rho}^2\sim \lambda^2M^2$ imagined by Kallosh and Linde in \cite{Kallosh:2007wm}. Such scenario remains able to produce observables acceptable with respect to the recent constraints published by Planck and BICEP2 \cite{Ade:2015tva, Ade:2015xua}. Observables obtained within this context are depicted in Fig. \ref{obs_Planck} for various values of the parameter $M$.

Yet, one should stress the fact that for low values of the parameter $M$, which could be taken for instance to be of order $M\sim$ GUT scale, the initial potential \eqref{GeneralPotential} can be minimized on the direction $\phi_1=\phi_2=0$ reproducing a model similar to a standard hybrid inflation arising before the GUT phase transition, similar to the scenario of \cite{Binetruy:1996xj}. In this framework, the inflaton is $|S|=(s^2+\sigma^2)^{1/2}$ and the effective inflation tree level potential is, at the fourth order in the sub-planckian vev of $S$, given by
\bea
V_0=
\frac{1}{8} \lambda ^2 M^4 \left[8+16 \zeta  \tilde{S}^2+ \left(32 \zeta ^2+14 \zeta +1\right) \tilde{S}^4\right]\,,
\eea
where we have normalized $|S|$ to $\tilde{S}= \sqrt{2} |S|$. However, for this scenario to be viable, and similarly to the hybrid inflation case, $\zeta$ must take negative values which is in a sense less motivated if one wants to interpret the quartic corrections to the K\"ahler as arising from loops \cite{Buchmuller:2014pla}.
\section{Adding Fayet Illiopoulos Terms}\label{sec:model2}
{In this section we study the influence of adding a constant Fayet Illiopoulos term}. As we will see this can perturb the inflation dynamics and will be constrained by observations.

The D-term potential is given by
\bea
V_D=\frac{g^2}{2}
\left(
|\phi_1|^2- |\phi_2|^2+\xi
\right)^2 \,,
\eea
and the SUSY minimum turns out to be
\begin{equation}\label{GlobalFIMin}
 |\phi_{1,2}|^2=\frac{\mp\xi+\sqrt{\xi^2+4M^4}}{2}\,.
\end{equation}

In such a scenario, the fields $s$ and $\sigma$ remain stabilized at zero. However, turning on the $\xi$ parameter will, contrarily to the previous section shifts the vev of $\alpha$ and $\beta$ to non-zero values. In order to understand clearly how fields stabilize in this case, one can thus define the directions in a polar representation as follows
\begin{eqnarray}
 \phi_1+\bar \phi_2&=&\left(R+R_0\right)e^{i\frac{\theta_1}{R_0}}\,,\\
\phi_1 - \bar \phi_2&=&\left(\rho+\rho_0\right)e^{i\frac{\theta_2}{\rho_0}}\,,
 \end{eqnarray}
$R_0$ and $\rho_0$ being the vacuum expectation values of the radial fields at the minimum. In this way, the goldstone boson arising from the breaking of the U(1) symmetry can be easily visualised in the combinations
 \bea
  \theta_1&=& (\kappa-\tau)\frac{\sqrt{\rho_0^2+R_0^2}}{\rho_0}\,,\\
  \theta_2&=& (\kappa+\tau)\frac{\sqrt{\rho_0^2+R_0^2}}{R_0}\,.
  \eea
Indeed the axion $\kappa$ does not appear in the potential expression and can be removed from the theory by a gauge transformation in a similar fashion that we did in Appendix \ref{unitarygauge}. Solving the equations of minimization $\partial_R V =0$ and $\partial_{\tau} V =0$ at the minimum imposes the condition
  \bea\label{phase}
  \langle\tau\rangle&=&\pi \frac{ \rho_0 R_0}{2 \sqrt{\rho_0^2+R_0^2}}\,,\\
  e^{R_0^2}&=&-\frac{4 g^2 R_0 (\rho_0 R_0-\xi )}{\lambda ^2 \rho_0 \left(-4 M^2+\rho_0^2-R_0^2\right)}\,,
  \eea
  and the condition $\partial_{\rho}V=0$ reads
  \be
 0= \frac{g^2 \left(R_0^2 \left(4 M^2-\rho_0^2+2\right)+2 \rho_0^2+R_0^4\right) (\rho_0 R_0-\xi )}{2 \rho_0}\,.
  \ee

 Solving both equations provides the exact vevs for $\rho$ and $R$
 \bea
  \langle\rho\rangle&\equiv&\rho_0=\left(2M^2+\sqrt{\xi^2+4M^4}\right)^{1/2}\,,\\
  \langle R\rangle&\equiv&R_0=\frac{\xi}{\left(2M^2+\sqrt{\xi^2+4M^4}\right)^{1/2}}\,.
  \eea
  The parameter $\xi$ being assumed to take values smaller than $M_p^2$, then the vevs shifting correspond to corrections of order $\mathcal O\left(\frac{\xi}{2M}\right)$ which will be used in what follows as the appropriate expansion parameter. Using the latter formalism makes appear clearly the Goldstone boson of the theory as an exactly massless field. However, the stabilization of the phase $\tau$ \eqref{phase}  renders the use of the basis $(\phi_1\pm\bar \phi_2)$  more appropriate to describe the scalar masses and the effective potential. Defining the basis
  \begin{equation}
  \phi_1 \pm \bar \phi_2 \equiv \alpha_{\pm} + i \beta_{\pm}\,,
  \end{equation}
the true supersymmetric vacuum reads
   \bea
 \langle s\rangle&=&\langle\sigma\rangle=\langle\beta_+\rangle=\langle\beta_-\rangle=0\,,\\
  \langle\alpha_-\rangle&=&\left(2M^2+\sqrt{\xi^2+4M^4}\right)^{1/2}\,,\\
  \langle \alpha_+\rangle&=&-\frac{\xi}{\left(2M^2+\sqrt{\xi^2+4M^4}\right)^{1/2}}\,.
  \eea

 As in the previous section, one can produce an inflationary scenario by letting the field $\alpha_-$ slow-roll during 60 e-folds while the other heavy fields stabilize quickly to their respective vacuum expectation values.

 Scalar masses at the end of inflation are given by
  \begin{eqnarray}
 m_s^2&=&m_{\sigma}^2=4M^2 \lambda ^2 \left(1+\frac{\xi ^2}{4M^2}+\dots\right)\,,\\
 m_{\alpha_-} ^2&=&m_{\beta_+}^2=\lambda ^2 \left(2 M^2 \left(1+\frac{\xi ^2}{4M^2}\right)+\frac{\xi ^2}{4M^2}+\dots\right)\,,\\
 m_{\alpha_+}^2&=&2 g^2 \left(2 M^2+\frac{\xi ^2}{4M^2}+\dots\right)\,,~~m_{\beta_-}^2=0\,.
 \end{eqnarray}
  where dots stand for higher order terms in $\mathcal O\left(\frac{1}{M^2}\frac{\xi^2}{4M^2}\right)$ or $ \mathcal O\left(\frac{\xi}{2M}\right)^4$  and $\beta_-$ main component is the goldstone boson which is hence an unphysical degree of freedom.

 During Inflation -- meaning that one allows $\alpha_-$ to take values away from its vev -- the field ${\alpha_+}$ is slightly shifted from its minimum and in the small vev approximation, one can obtain the values
 \begin{eqnarray}
 \langle\beta_+\rangle_{inf}&\simeq& 0\,,\\
 \langle\alpha_+\rangle_{inf}&\simeq& -\frac{8 g^2 \xi  {\alpha_-}}{8 g^2 {\alpha_-} ^2 +\lambda ^2 {\alpha_-} ^4-2 \lambda ^2 {\alpha_-} ^2+16 \lambda ^2 M^4-8 \lambda ^2 M^2 {\alpha_-} ^2+8 \lambda ^2 M^2}\,.
\end{eqnarray}
Under the assumption $\xi\ll 2M$ -- which is reasonable since we impose $M\gg1$ -- such values of the fields are indeed highly suppressed. Minimizing the potential with respect to $s,\sigma,{\alpha_+}$ and $\beta_+$ hence provides the effective inflation potential
\bea\label{pot2}
V_{inf}({\alpha_-})&=&\frac{1}{16} \left(8 g^2 \left(\xi -{\alpha_-} A\right)^2+\lambda ^2 e^{A^2} \left( \left(A^2-{\alpha_-} ^2\right)+4 M^2\right)^2\right)\,,\nonumber\\
A({\alpha_-})&\equiv&\frac{8 g^2 \xi  {\alpha_-}}{{\alpha_-} ^2 \left(8 g^2+\lambda ^2 \left({\alpha_-} ^2-2\right)\right)+16 \lambda ^2 M^4-8 \lambda ^2 M^2 \left({\alpha_-} ^2-1\right)}\,.
\eea
%
 During Inflation, the tree level mass matrix gets, from the FI term, off diagonal terms that have to be removed by diagonalization of the blocks
\bea\label{block1}
(m^{\text{inf}}_{\beta_-,\beta_+})^2=\left(
\begin{array}{cc}
 \frac{\lambda ^2}{4}({\alpha_-} ^2-4M^2)& g^2 \xi  \\
 g^2 \xi  &    \frac{\lambda ^2}{8}  \left({\alpha_-} ^4 + {\alpha_-} ^2 (2 - 8 M^2) + 8 M^2(1 + 2 M^2)\right)\\
\end{array}
\right)
\,,
\eea
and
\bea \label{block2}
(m^{\text{inf}}_{{\alpha_+},{\alpha_-}})^2=\left(
\begin{array}{cc}
 \frac{\lambda ^2 {\alpha_-} ^4}{8}+\left(g^2-\frac{1}{4} \left(4 M^2+1\right) \lambda ^2\right) {\alpha_-} ^2+M^2 \left(2 M^2+1\right) \lambda ^2 & -\frac{\xi g^2 ({\alpha_-}+6 M)}{4M}  \\
 -\frac{\xi g^2 ({\alpha_-}+6 M)}{4M} & \frac{1}{4} \left(3 {\alpha_-} ^2-4 M^2\right) \lambda ^2 \\
\end{array}
\right)
\,,\n
\eea
while the S field still gets masses
\bea
(m^{\text{inf}}_s)^2&=&(m^{^{\text{inf}}}_{\sigma})^2=\frac{\lambda ^2}{2}  \left(\zeta  \left({\alpha_-} ^2-4 M^2\right)^2+2 {\alpha_-} ^2\right) \,.
\eea
Eigenvalues of the latter blocks are, at leading order in $\frac{\xi}{2M}$ and $M\gg1$, given by
\begin{eqnarray}\label{massesInfFI}
(m^{^{\text{inf}}}_{{\alpha_+}})^2&=&{\alpha_-} ^2 g^2+\frac{\lambda ^2}{8}\left({\alpha_-} ^4-2 {\alpha_-} ^2 \left(4 M^2+1\right)+8 \left(2 M^4+M^2\right)\right)\n
&&+\frac{9 g^4}{2 \lambda ^2 M^2}\frac{\xi ^2}{4M^2}+\dots\,,\\
 (m^{^{\text{inf}}}_{{\alpha_-}})^2&=&\frac{1}{4} \lambda ^2 \left(3 {\alpha_-} ^2-4 M^2\right)-\frac{9 g^4}{2 \lambda ^2 M^2}\frac{\xi ^2}{4M^2}+\dots\,,\\
  (m^{^{\text{inf}}}_{\beta_-})^2&=&\frac{1}{4} \lambda ^2 \left({\alpha_-} ^2-4 M^2\right)-\frac{ g^4}{2 \lambda ^2 M^2}\frac{\xi ^2}{4M^2}+\dots\,,\\
 (m^{^{\text{inf}}}_{\beta_+})^2&=&\frac{\lambda ^2}{8}\left({\alpha_-} ^4+{\alpha_-} ^2 \left(2-8 M^2\right)+8 \left(2 M^4+M^2\right)\right)+\frac{ g^4}{2 \lambda ^2 M^2}\frac{\xi ^2}{4M^2}+\dots\,.
 \end{eqnarray}
 {These masses are} corresponding to  {those} we got without FI terms  {in} (\ref{masses}){, in addition to} corrections of order $\mathcal O\left(\frac{\xi}{2M}\right)^2$. Again, the field $\beta_-$ is mainly constituted of the Goldstone boson $\kappa$, and turns out to be an unphysical degree of freedom  {and hence is} removable from the theory by a gauge transformation.
Among other masses, and for values of $\zeta \gtrsim \frac{1}{24}$, only ${\alpha_-}$ has a mass smaller than the Hubble scale and can be the single field driving the Inflation process.

Numerical simulation of the observables in the case $\xi \lesssim M_p \ll M$ provide results essentially identical to the case without FI terms. Indeed corrections of the dynamics have been shown to appear through the expansion parameter $\xi/M$. If one wishes to explore regions of the parameter space where $\xi \gg 2M$ and in particular where $M$ is taken to be sub-planckian, one has to solve numerically the equations of motion in order to determine the -- Inflaton dependant -- vacuum expectation value of ${\alpha_+}$. One can in this case release 60 e-folds of Inflation while the Inflaton rolls down from transplanckian values, but such scenarios lead to observables where typically
\begin{equation}
 n_s \sim 0.965 \text{~~~and ~~~}r\gtrsim 0.2\,,
\end{equation}
which is excluded to high significance by Planck and Bicep II measurements\cite{Ade:2015xua, Ade:2015tva}. Note that the parameter space is far from being free of constraint since a too high value of the product $g \xi$ (typically when $g^2 \xi^2 \gtrsim H^2$) can violate the COBE normalization. Furthermore, the shift symmetry imposed to protect the inflaton mass from dangerous corrections is in this case violated and the $\eta$--problem re-appears since the mixing in the mass matrices (\ref{block1}, \ref{block2}) is no longer negligible.

\paragraph{Comment on local R-symmetry}

 As studied in full details in \cite{Binetruy:2004hh}, once going to supergravity with a constant FI term, one has to redefine charges such that the superpotential transforms correctly under the induced local R-symmetry. In this case, the invariance of the superpotential (Eq. \ref{SPot1}) under $U(1)$ gauge symmetry imposes \cite{Binetruy:2004hh} that
 \begin{equation}
\delta W = i \sum_{i} g q_i \phi_i \partial^i W = i\frac{g\xi}{M_p^2}W\,,
 \end{equation}
where
\begin{equation}\label{charge-constraint}
q_i\,=\,Q_i +  \frac{a_i}{2} \frac{\xi}{M_p^2}\,, ~~~~~~~~~\sum_i a_i\,=\, 2\,,
\end{equation}
with $a_i$ are the R--symmetry charges and $Q_i$ are $U(1)$ charges of the chiral superfields $S,\,\phi_1,\,\phi_2$ in globally SUSY limit, {\it i.e.}, $Q_i$ have the following assignments: $Q_S=0,~ Q_1=1,~ Q_2=-1$. Such condition imposes that
\begin{equation}
 q_S = \frac{\xi}{M_p^2}\text{~~~and~~~}q_1+q_2 =0\,,
\end{equation}
with
\begin{eqnarray}
a_S=2 \text{~~~and~~~}a_1=-a_2\,,
\end{eqnarray}
which actually coincides with the R-charges in the global limit. Hence,
the D-term potential is then given by
\bea
V_D=\frac{g^2}{2}
\left(q_S\,|S|^2+
q_1\,|\phi_1|^2+ q_2\,|\phi_2|^2+\xi
\right)^2
\eea
As a matter of fact, one could expect from such modification of the D-term potential to perturb the inflationary scenario. Nevertheless it is not the case. Indeed, the presence of the additional piece in the D-term potential providing more mass to the field $S$, and the charges $q_1$ and $q_2$ being still of opposite charges still imposes a stabilization of the fields around the global minima (\ref{GlobalFIMin}), where the FI term $\xi$ is rescaled as $\xi\,\to\, \xi/q$ with $q=1+\xi /(2M_p^2)$.

Correspondingly, $ \alpha_{\pm}, \beta_{\pm}$ at the minimum will be slightly
shifted\footnote{We assume here that $\xi < M_p$, which constitutes a reasonable assumption from a string theory point of view. Note in particular that too high values of $\xi$ require very small values of the parameter $g$ for reasons mentioned above, which would also be somehow unnatural. Otherwise $\xi$ should be extremely small compared to the Planck scale.}
\bea
 \langle s\rangle&=&\langle\sigma\rangle=\langle\beta_+\rangle=\langle\beta_-\rangle=0\,,\\
  \langle\alpha_-\rangle&=&\left(2M^2+\sqrt{\xi^2+4M^4}\right)^{1/2}\, - \, \frac{1}{4 M}\frac{\xi^2}{ M_p^2} \,+\, {\cal O}(\frac{\xi^4}{ M_p^4})\,,\\
  \langle \alpha_+\rangle&=&-\frac{\xi}{\left(2M^2+\sqrt{\xi^2+4M^4}\right)^{1/2}}\,+\, \frac{1}{16 M^3}\frac{\xi^3}{ M_p^2} \,+\, {\cal O}(\frac{\xi^3}{ M_p^4})\,.
\eea

Masses of the fields remains essentially unchanged (up to the rescaling mentioned above), whereas as we just mentioned the mass of the field $S$ gets corrections
\bea
(m^{\text{inf}}_s)^2&=&(m^{^{\text{inf}}}_{\sigma})^2=\frac{\lambda ^2}{2}  \left(\zeta  \left({\alpha_-} ^2-4 M^2\right)^2+2 {\alpha_-} ^2\right)+2q_S g^2 \xi \,.
\eea

 As an interesting point, one could yet notice that the presence of a new quadratic term in $S$ in the scalar potential may allow to get rid of the quartic term introduced in the K\"ahler in order to stabilize the latter. Indeed, for ${g^2\xi^2}\gtrsim H^2$ such mass term could play such role and provide a mass for $S$ larger than the Hubble scale. Yet, as we discussed previously such region of the parameter space remains excluded since diagonalization of the mass matrix would destroy the shift symmetry and re-introduce an $\eta$-problem.

To put the above results in a nut-shell, one can conclude that the effects of introducing a Fayet-Illiopoulos term would become relevant when the parameter $\xi$ becomes of order $2M\gg1$, which would be somehow difficult to motivate. In the reasonable case where $\xi\ll 2M$, effects of the FI term appear furthermore as negligible perturbations of the inflation potential \eqref{pot2} and have no significant consequences on the cosmological observables.

\section{Reheating After inflation}
\label{sec:reheat}
We now turn to study the reheating  process after inflation. In this process, the known matter is created through the oscillation and the decay of the inflaton field. In standard supersymmetric hybrid inflation models, the inflaton is a pure singlet, therefore, it is tempting to assume an unknown non-renormalizable coupling between the inflaton and right-handed neutrinos, which is generally suppressed by Planck scale. In such case the inflaton can decay to pairs of right-handed neutrinos which can generate the observed baryon asymmetry of the universe via leptogenesis.

In our model, the inflaton is charged under a $U(1)$ gauge symmetry, and may thus naturally decay into particles charged under this symmetry or any other $U(1)'$ that has the usual kinetic mixing with our $U(1)$.  One may assume that the MSSM Higgs fields $H_u$ and $H_d$ are charged under this $U(1)$ and hence the reheating process may occur if the inflaton decays to Higgs fields via the mediation of the $U(1)$ gauge boson. However, any direct coupling like $\rho H_u H_d$ will generate a very large $\mu$-term, which is phenomenologically unacceptable. Also the possible decay of $\rho$ to two gauge bosons that decay into $H_u$ and $H_d$ (and eventually SM quarks through Yukawa interactions) will be highly suppressed and can not account for the required decay width.

\begin{figure}[ht]
\begin{center}
\includegraphics[width=0.6\linewidth]{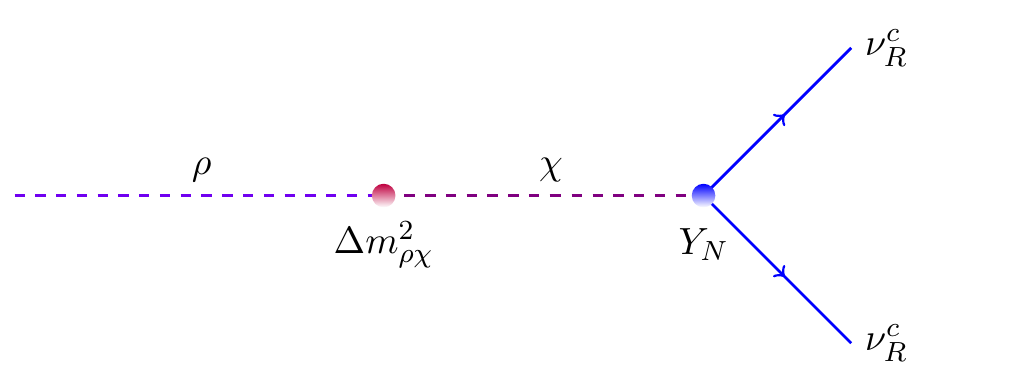}
\end{center}
\caption{\label{eff_coupling}{\footnotesize Process inducing a decay of the inflaton into the right-handed neutrinos $\nu_R^c$. The diagram involves a mixing $\Delta m^2_{\rho \chi} = \frac{\tilde g\ g_{BL}}{2} v_{B-L} \langle \rho \rangle$ between the inflaton $\rho$ and the $B-L$ higgs field $\chi$ whose value arises from kinetic mixing of the two $U(1)$'s (see Appendix \ref{Inflationary BLSSM}).}}
\end{figure}

It is clear that our $U(1)$ can not be the $U(1)_{B-L}$ of right-handed neutrinos since it would be broken at a very large scale close to the Planck scale and hence be inconsistent with the expected neutrino masses \cite{Ade:2015xua}. Therefore, we consider a $U(1)_{B-L}$ as an extra symmetry that has a kinetic mixing with our $U(1)$. We assume in what follows the $B-L$ symmetry to be broken at a scale $v_{B-L}\sim 10^{13} \mathrm{GeV}$ such that the associated scalar and gauge fields get masses after the inflation ends. The possible kinetic mixing between our $U(1)$ and the $U(1)_{B-L}$ would induce a mixing between their gauge bosons. In addition, it was shown in \cite{Abdallah:2014fra, O'Leary:2011yq} that in the context of the BLSSM (a $U(1)_{B-L}$ extension of the MSSM) the kinetic mixing between $U(1)_Y$ and $U(1)_{B-L}$ generates a mixing between the standard model Higgs and the $B-L$ scalar field $\chi$. Also in our case one may obtain a similar mixing between the inflaton field and the $B-L $ scalar field $\chi$.  As shown in Appendix \ref{Inflationary BLSSM}, the $\rho-\chi$ mixing is given by
\begin{equation}
\delta_{\rho\chi} = \frac{\tilde g\, g_{BL}\, v_{B-L}\, \langle \rho \rangle }{2(m_{\rho}^2-m_{\chi}^2)}\,,
\end{equation}
where $\tilde g$ and $g_{B-L}$ are the kinetic mixing parameter and $U(1)_{B_L}$ gauge coupling respectively. Such a mixing generates a coupling between the inflaton and right-handed neutrinos $\nu_R^c$, as depicted in Fig. \ref{eff_coupling}.

Thus, the effective coupling between the inflaton  and right-handed neutrinos, $Y_{eff}$, can be written as
\begin{equation}
 Y_{eff} = \delta_{\rho \chi} Y_N\,,
\end{equation}
where $Y_N$ is the Yukawa coupling of scalar field $\chi$ to the right-handed neutrino $\nu_R^c$.

The inflaton mass is fixed by the COBE normalization \cite{Smoot:1992td} $ m_{\rho} \sim \lambda M \sim {\cal O} (10^{13}) \mathrm{GeV}$ and we assume here that the scale of spontaneous breaking of the B-L gauge symmetry matches the inflaton mass scale $v_{B-L}\sim m_{\rho}$. The mass of $m_{\chi}$ is however not protected by the $B-L$ symmetry breaking scale and can thus be larger than the latter (even of order GUT scale due to possible mixings with other GUT scale scalars). The mixing $Y_{eff}$ turns hence out to be of order
\be
Y_{eff} \simeq  \tilde{g}\, g_{B-L} \left(\frac{v_{B-L}}{m_\chi}\right) \left(\frac{M}{m_\chi}\right)\, Y_N.
\ee
For $m_\chi \sim {\cal O}(10^{16})$ GeV, one thus finds $Y_{eff} \simeq   \tilde{g}\,g_{B-L}\, Y_N$. The reheating temperature $T_R$ is given by~\cite{Lazarides:2001zd, Lazarides:1996dv}
\bea
T_R \approx \frac{ (8\pi)^{1/4}}{7}\left( \Gamma\,  M_p\right)^{1/2},
\eea
where $\Gamma$ is the total decay width of the inflaton field
\be
\Gamma = \Gamma_{\rho \to \nu_R^c\nu_R^c} + \Gamma_{\rho\to \tilde\nu_R^c {}^* \tilde{\nu}_R^c}\,,
\ee
and  $M_p$ is the reduced Planck mass. The interactions of $\rho$, $\nu_R^c$ and $\tilde{\nu}_R^c$ are given by
\begin{equation}
 \mathcal L \supset Y_{eff}\left( \rho \ \nu_R^c  \nu_R^c + M_{\nu_R} \ \rho\ \tilde\nu_R^c{}^* \tilde\nu_R^c\right)\,,
\end{equation}
with $M_{\nu_R}\sim Y_N v_{B-L}$ is the supersymmetric mass of the right-handed neutrino supermultiplet. The decay rates are then
\bea\label{dec1}
\Gamma_{\rho \to \nu_R^c\nu_R^c} &=& \frac{ Y_{eff}^2 }{16 \pi}\frac{\left( m_{\rho}^2-4  M_{\nu_R}^2\right)^{3/2}}{ m_{\rho}^2} \,,\\
\Gamma_{\rho\to \tilde\nu_R^c{}^*\tilde{\nu}_R^c} &=& \frac{ Y_{eff}^2 M_{\nu_R}^2}{8 \pi}\frac{\sqrt{m_{\rho}^2 - 4 M_{\nu_R}^2}}{ m_{\rho}^2}\,.
\eea
Therefore, $T_R$ takes the form
\bea\label{TR}
T_R &=& \frac{(8\pi)^{1/4}M_p^{1/2} Y_{eff}\left( m_{\rho}^2-4  M_{\nu_R}^2\right)^{1/4}}{28\sqrt{\pi}}\left[1- 2\frac{M_{\nu_R}^2}{m_{\rho}^2}\right]^{1/2}\,,\\
&\approx&\frac{(8\pi)^{1/4}(M_p\ m_{\rho})^{1/2} }{28\sqrt{\pi}}Y_{eff}\,.
\eea
For the aforementioned values of masses and $B-L$ coupling one can show that the reheating temperature is of order $T_R \sim 10^{14} \, \tilde{g}\,g_{B-L}\,Y_N$ GeV. Thus, for $g_{B-L} \sim {\cal O}(0.1)$, $Y_N\sim 10^{-2}$ and  $\tilde{g} \sim 10^{-2}$ one gets $T_R\sim 10^{9}$~GeV, which is consistent with the cosmological constraints \cite{Ellis:1983ew,Khlopov:1984pf, Linde:2005ht,Kawasaki:1994af,Moroi:1995fs}.

Finally, it is  worth recalling that the inflaton field has mass $\sim {\cal O}(10^{13})$ GeV and that all the other scalar fields are much heavier, so that they are stabilized in the global minimum during inflation. Although some of them (heavy components of the fields $\phi_{1,2}$) are charged under $U(1)$ and could in principle produce lighter particles much before the inflation ends, their energy density is then diluted during inflation due to the fast expansion of spacetime such that the latter particles play no role in the inflationary scenario.

Before concluding, it is important to note that our analysis can be generalised for a supergravity framework with a general gauge group $G$ (say $SO(10)$ or flipped $SU(5)$) instead of $U(1)$, especially as emphasized above  where a viable inflationary scenario is obtained without FI term. In this case, the charged fields $\phi_1$ and $\phi_2$ will be chosen as fundamental and anti-fundamental multiplets of the gauge group $G$. The vevs of these scalars will break $G$ down to a subgroup $H$ that contains the SM gauge group $SU(3)_c \times SU(2)_L \times U(1)_Y$. Thus inflation will be associated to the high scale phase transition and gauge symmetry breaking.

Finally, we emphasize that in many models, where inflation is  embedded in  grand unified scenarios \cite{Carpenter:2014saa, Ellis:2014xda, Pallis:2014xva,Watari:2004xh, Connors:1988yx},  the associated gauge group is usually broken by the waterfall of charged spectator fields at the end of inflation. Therefore, a  dangerous induced production of topological defects (such as monopoles, cosmic strings, etc.) has to be kept under control \cite{Pallis:2013dxa, Ade:2013xla, Rocher:2004et, Buchmuller:2012wn}.  A way to overcome this problem was suggested in what is known as smooth and shifted hybrid inflation \cite{Lazarides:1995vr, Jeannerot:2000sv, Jeannerot:2002wt, Jeannerot:2001qu}, where the gauge symmetry is broken by spectator fields along the inflationary path of a singlet field. As advocated in the introduction, in our scenario the $U(1)$ symmetry, or GUT gauge group in general, is broken along the inflation trajectory, hence  any produced cosmic string will be diluted during inflation. In this respect, charged field inflation seems to be favoured.
\section{Conclusions}
In this paper we studied a single field inflation scenario within supergravity with shift and abelian $U(1)$ symmetry. We have considered the case where the inflaton is charged under anomalous and non-anomalous $U(1)$. We have shown that an effective potential of the {\em new inflation} type is obtained in both cases, which allows to release 60 e-folds of large field inflation with consistent spectral index and tensor to scalar ratio with the recent Planck and BICEP II results. We have also analysed the reheating after  inflation due to the decay of the inflaton. We emphasized that one can generate natural couplings between the inflation and right-handed (s)neutrinos, through the kinetic mixing between our $U(1)$ and $U(1)_{B-L}$.  With intermediate $B-L$ symmetry breaking scale, we showed that the reheating temperature can be of order $10^9$ GeV, consistently with usual cosmological constraints.
\section*{Acknowledgments}
S.K. acknowledge support by the French Institute of  Egypt and would like to thank the CPhT at Ecole Polytechnique for hospitality.
The work of SK and AM is partially supported by the STDF project 13858, the ICTP grant AC-80 and the  European Union FP7  ITN INVISIBLES (Marie Curie Actions, PITN- GA-2011- 289442).
 LH wants to thank the DESY Theory group of Hamburg for its hospitality and funding during the realization of this work. The authors are very grateful to Emilian Dudas for his precious comments and suggestions.

\appendix

\section{Elimination of the Goldstone Boson}
\label{unitarygauge}
Expanding the gauge Lagrangian~(\ref{gaugelag}) after shifting the fields around the minimum
\begin{eqnarray}\label{gaugelag1}
 \mathcal L_G\equiv K_{I\bar J}D_{\mu}\phi^I \overline{D^{\mu}\phi^{J}} &=& |(\partial_{\mu}-igA_{\mu})\phi_1|^2+|(\partial_{\mu}+igA_{\mu})\phi_2|^2\,,\\
 &=&|(\partial_{\mu}-igA_{\mu})(M e^{i\theta/2M})|^2+|(\partial_{\mu}+igA_{\mu})(-Me^{-i\theta/2M})|^2\,,
\end{eqnarray}
hence, one finds a mixing term of the form
\bea\label{unphys}
-2g M A^\mu \partial_\mu \theta\,.
\eea
This term is unphysical and we can get rid of it by fixing the gauge to the unitary gauge and accordingly the Nambu Goldstone boson $\theta$ will be eaten by the gauge
field to acquire its physical mass according to the Higgs mechanism. Now let us discuss and the gauge fixing.

Referring to the field redefinitions~(\ref{def}), one can write the gauge Lagrangian (\ref{gaugelag1}) in terms of $\rho, \theta, \alpha, \beta$.
\bea
\phi_1+\bar \phi_2 &\equiv& \alpha+i \beta \\ \nonumber
\phi_2-\bar \phi_2 &\equiv& \rho \ e^{i \theta / 2M} \,,
\eea
While fields oscillate around their minima, one can define the unitary gauge by setting
\bea\label{gaugefixing}
B^\mu &\equiv&  A^\mu -\frac{1}{2Mg}\partial_\mu \theta\\ \nonumber
 \tilde \alpha+i \tilde \beta&\equiv& e^{-i\theta/(2M)} (\alpha+i \beta)\,.
\eea
In this respect, the $\theta$ degree of freedom will disappear and be absorbed so that the massless $U(1)$ gauge field is realized as a massive field.
The gauge lagrangian (\ref{gaugelag1}), under these transformations, becomes
 \bea
 \mathcal L_G&=&\frac{(\partial\tilde\alpha_1)^2+(\partial\tilde\beta_2)^2+(\partial\rho)^2}{2}+\frac{B_{\mu}^2 g^2}{2}  \left(\tilde\alpha_1^2+\tilde\beta_2^2+\rho^2\right)+g B_{\mu} (\tilde\beta_2 \partial_{\mu}\tilde\alpha_1 - \tilde\alpha_1 \partial_{\mu}\tilde\beta_2 )\,.
 \eea

 It hence appears after expanding the inflaton vev around its vev $\langle\rho\rangle=2M$ that the gauge boson gets in the ground state a tree level mass $m_B^2 = 2 g^2 M^2$. Note that this elimination of the goldstone boson $\theta$ is valid after and during inflation, as stated in section \ref{sec:model1}.
\section{Inflationary BLSSM}
\label{Inflationary BLSSM}

In this appendix we aim to give a rigorous formulation of the microscopic model proposed in section \ref{sec:reheat} to provide a decay channel for the inflaton. The discussion is based on the works realized in \cite{O'Leary:2011yq, Abdallah:2014fra}.

We assume that the theory is described by two $U(1)$ symmetries : one under which the inflaton is charged, and being the B-L symmetry used for leptogenesis. The charges of the fields is summed up in Tab. \ref{BLSSM}.

\begin{table}
\begin{tabular}{c | c c}
 &$U(1)_{inf}$&$U(1)_{B-L}$\\\hline \hline
 $\phi_1$ & +1 & 0\\
 $\phi_2$ & -1 & 0 \\
 $\chi_1$ & 0 & +1\\
 $\chi_2$ & 0 & -1
\end{tabular}
\caption{\label{BLSSM}{\footnotesize Charges of the fields involved in the mixing between our $U(1)$ and the B-L $U(1)$ symmetry.}}
\end{table}

Recall that in our model the inflaton $\rho$ is the real component (for a convenient choice of $\theta$) of the combination

\begin{equation}
 \phi_1-\bar\phi_2 \equiv \alpha_2 + i  \beta_1=\rho + 2M \equiv \rho + v_{inf}\,.
\end{equation}
On the other hand we showed that $\langle \phi_1 + \bar\phi_2\rangle = \langle \alpha_1+ i\beta_2\rangle=0$ such that we can generically write the initial fields $\phi_i$ under the form
\bea
\phi_1 &=& \frac{1}{\sqrt{2}}\left(v_1^{inf} + \rho_1 + i\ \varphi_1\right)\,,\\
\phi_2 &=& \frac{1}{\sqrt{2}}\left(v_2^{inf} + \rho_2 + i\ \varphi_2\right)\,,
\eea
where $v_1^{inf} = -v_2^{inf} = v_{inf} / \sqrt{2}$. The fields $\chi_{1,2}$ breaking the $U(1)_{B-L}$ symmetry can be written in the same fashion than in \cite{Abdallah:2014fra}
\bea
\chi_1 &=& \frac{1}{\sqrt{2}}\left(v_1 + l_1 + i\ m_1\right)\,,\\
\chi_2 &=& \frac{1}{\sqrt{2}}\left(v_2 + l_2 + i\ m_2\right)\,.
\eea
We now allow a kinetic mixing the gauge fields of two $U(1)$'s. According to \cite{Abdallah:2014fra, O'Leary:2011yq}, rotating the gauge fields in a basis $A^{inf}_{\mu} , A^{B-L}_{\mu}$ in which the kinetic mixing takes the form
\begin{equation}
 \begin{pmatrix}
                                                            g_{inf} & \tilde g \\
                                                            0 & g_{BL}
                                                           \end{pmatrix}\,,
\end{equation}
a mass matrix mixing block between the real scalar fields $(\rho_1, \rho_2 , \chi_1,\chi_2)$ is induced from writing the D-terms and can be written to be
\begin{equation}
 m^2_{\rho_{1,2} \chi_{1,2}} = \frac{1}{2} \tilde g g_{BL} \begin{pmatrix}
                                                            v_1^{inf}v_1 & -v_1^{inf}v_2 \\
                                                            -v_2^{inf}v_1 & v_2^{inf}v_2
                                                           \end{pmatrix}\,,
\end{equation}
which gives, after a rotation in the basis $\left(\rho\equiv\frac{1}{\sqrt{2}}(\rho_1-\rho_2), \tilde\rho\equiv\frac{1}{\sqrt{2}}(\rho_1+\rho_2) , \chi_1,\chi_2\right)$
\begin{equation}
 m^2_{\rho \tilde \rho \chi_{1,2}} = \frac{1}{2} \tilde g g_{BL} \begin{pmatrix}
                                                            v_{inf}v_1 & -v_{inf}v_2 \\
                                                            0 & 0
                                                           \end{pmatrix}\,.
\end{equation}

This writing of the mixing allows us, since the field $\chi_1$ is coupling to right handed neutrinos in a BLSSM setup via the superpotential term
\begin{equation}
 W \supset Y_N \chi_1 N_R^c N_R^c\,,
\end{equation}
to provide the inflaton a tree level decay, as studied in section \ref{sec:reheat}. Note that the vev $v_1$ is denoted by $v_{B-L}$ in the rest of the paper and that the field $\chi_1$ is denoted by $\chi$ for notation simplicity.

\bibliographystyle{utphys}   

\end{document}